# Equivalence of Fluctuation-Dissipation and Edwards' Temperature in Cyclically Sheared Granular Systems

Zhikun Zeng,[1] Shuyang Zhang,[1] Xu Zheng,[3] Chengjie Xia,[4] Walter Kob,[5,1] Ye Yuan,[1,*] and Yujie Wang[1,2,3,†]

[1]School of Physics and Astronomy, Shanghai Jiao Tong University, Shanghai 200240, China
[2]State Key Laboratory of Geohazard Prevention and Geoenvironment Protection, Chengdu University of Technology, Chengdu 610059, China
[3]Department of Physics, College of Mathematics and Physics, Chengdu University of Technology, Chengdu 610059, China
[4]School of Physics and Electronic Science, East China Normal University, Shanghai 200241, China
[5]Laboratoire Charles Coulomb, University of Montpellier and CNRS, 34095 Montpellier, France

Using particle trajectory data obtained from x-ray tomography, we determine two kinds of effective temperatures in a cyclically sheared granular system. The first one is obtained from the fluctuation-dissipation theorem which relates the diffusion and mobility of lighter tracer particles immersed in the system. The second is the Edwards compactivity defined via the packing volume fluctuations. We find robust excellent agreement between these two temperatures, independent of the type of the tracers, cyclic shear amplitudes, and particle surface roughness. We further elucidate that in granular systems the viscous-like drag force is due to the broken symmetry of the local contact geometry.

Understanding disordered materials far from thermal equilibrium is one of the biggest challenges in condensed matter physics [1]. Many principles of equilibrium statistical mechanics have proven to be still valid for characterizing the non-equilibrium behaviors of these systems, with the concept of effective temperature playing an important role, rather than the thermal bath temperature [2,3]. Various studies have shown that the effective temperature is useful for understanding the structural relaxation [4-7], plasticity [8,9], and rheological properties of various non-equilibrium disordered materials [10,11]. However, multiple approaches exist to define an effective temperature, and both the physical meaning and the relationship between these temperatures have so far remained elusive [3].

Granular matter, which is an achetypical disordered system ubiquitous in nature and engineering processes, also needs a statistical mechanics framework that allows the definition of an effective temperature [12]. It is found that under consecutive external driving, granular systems evolve into stationary packings with a mildly fluctuated packing fraction (or packing volume) irrespective of preparation history, reminiscent of the thermodynamic fluctuation of thermal system at constant temperature [13,14]. In the 1990s, Edwards and collaborators introduced a statistical mechanics framework for jammed granular packings based on this observation [15]. According to their approach, volume is postulated to be the quantity equivalent to energy in thermal systems. It is further conjectured that jammed packings of the same volume are equally probable, validated so far only in numerical simulations of frictionless packings [16-18]. Such an approach naturally leads to the definition of entropy as the logarithm of the number of all jammed states in the microcanonical ensemble, and a compactivity $\chi$ that plays the role of the conventional temperature can be defined accordingly [19]. Recent experiments have demonstrated the validity of Edwards volume ensemble in frictional jammed granular packings [20].

Another approach to define an effective temperature $T_{FD}$ is based on the fluctuation-dissipation relation (FDR), which has been widely investigated for granular materials in the stationary state [21-25]. In the FDR, $T_{FD}$ relates the diffusion and the mobility of a tracer particle over a period of time $t$:

$$k_B T_{FD} = \frac{\langle [r(t)-r(0)]^2 \rangle}{2\langle r(t)-r(0)\rangle/F}, \tag{1}$$

where $r$ is the position of the particle, and $F$ is a weak external force acting on the particle balanced by the drag force in the stationary directional moving regime. Simulations have shown that in frictionless jammed granular packings $T_{FD}$ is equal to $\chi$ [16,17], so that $\chi$ governs the transport properties of these systems. However, there have been no experimental studies investigating the relevance between $\chi$ and $T_{FD}$ for frictional granular systems so far, which is not only critical for the understanding of granular rheology, but also hampers the development of non-equilibrium statistical mechanics in a broader sense [26,27].

In this letter, we use x-ray tomography to determine the compactivity $\chi$ as well as the $T_{FD}$ of a three-dimensional granular system under quasistatic cyclic shear. In practice, we measure $\chi$ from the microscopic volume fluctuations within the disordered packings. Using hollow particles (HP) with different lower mass densities as compared to the background particles (BP), we measure $T_{FD}$ based on Eq. (1) by tracking both the HPs' directional motion due to buoyancy and diffusive motion. We find that $T_{FD}$ and $\chi$ agree with each other robustly for all the steady states investigated.

We 3D-print (ProJet MJP 2500 Plus, 0.032 mm resolution) the HPs and BPs of a same plastic material (VisiJet M2R-WT, $\rho = 1.12 \times 10^3$ kg/m$^3$) and diameter $d = 6$ mm. For both HPs and BPs, we prepare particles with two types of surface properties: a smooth surface (SS) and a bumpy surface (BS) realized by uniformly decorating its surface with 150 hemispheres of radius 0.04$d$, which mimics a particle with very large surface friction. The HPs are lighter than the solid BPs, and their mass difference is $\Delta m = m_0 - m$, where $m$ and $m_0 = 0.1263 \pm 0.0001$ g are the respective masses of HPs and BPs. We prepare 13 types of HPs for the SS systems with $\Delta m / m_0 \in [0.017, 0.484]$, and 5 types of HPs for the BS systems with $\Delta m / m_0 \in [0.031, 0.500]$.

Figure 1(a) is the schematic of our experiment setup. The shear cell has a cuboid shape with a size of 120 mm ($x$) × 120 mm ($y$) × 140 mm ($z$), where the bottom and side walls are rendered rough by gluing a layer of hemisphere particles at random positions to prevent crystallization. The cyclic shear is generated by a step motor attached to the bottom plate of the shear cell. See Ref. [28] for further details of the setup. For each measurement, a packing in the shear cell contains ~12,000 particles of a height ~$22d$. For better statistics, we manually immerse 100 HPs with a specific $\Delta m$ uniformly in the packing within a height interval of 8~$12d$ initially. The HPs are spaced sufficiently apart from each other to avoid any collective effect. Furthermore, to investigate the influence of pressure $p$ on the dynamics of the system, we perform experiments on the packings either with a free upper surface or covered by a lid of mass $M$ of 1.95 or 3.85 kg. Correspondingly, the average imposed pressure at the HPs' height is directly measured to be $p$ = 0.35 ± 0.07, 1.67 ± 0.10, and 2.86 ± 0.17 × $10^3$ Pa (see Supplemental Material [29] for more details).

The cyclic shear is applied with a strain rate of $\dot{\gamma} = 0.33$ s$^{-1}$ and different strain amplitudes $\gamma$ =0.03, 0.05, 0.08, 0.12 and 0.20. The inertial number is on the order of $10^{-4}$ for all cases investigated, so that the shear is quasistatic [30]. Given each different value of $\gamma$, we apply hundreds of shear cycles to the initial packing until a steady state is reached and a large $\gamma$ leads to a lower steady state packing fraction. After the system reaches a steady state, we obtain its packing structure via a medical CT scanner (UEG Medical Group Ltd., 0.2 mm spatial resolution) after every 10 shear cycles. A total of 200 CT scans for the SS systems and 150 CT scans for the BS systems are taken for each $\gamma$ and $\Delta m$. To further improve the statistics, we repeat the measurements for each type of HPs 3 or 4 times. Following the image processing procedures and tracking algorithms of our previous study [28], one can obtain the centroid coordinates with an error less than $3 \times 10^{-3} d$ of each particle. Only particles that are at least $3d$ away from the boundary of the shear cell are analyzed in the following.

For the system of SS particles and a free upper surface, the trajectory of a HP ($\Delta m / m_0 = 0.376$) under cyclic shear ($\gamma = 0.05$) is shown in the inset of Fig. 1(b). It is clear that a HP displays both diffusive motion and vertical directional motion under cyclic shear, analogous to the dynamics of a Brownian particle subject to a directional force in a thermal liquid. The one-dimensional diffusive motion of a thermal Brownian particle can be characterized by the mean square displacement (MSD), $\langle [z(t) - z(0)]^2 \rangle = 2Dt$, where $D$ is the self-diffusion coefficient. If an external force $F$ is applied to the particle, the particle will experience a viscous drag force from the liquid and the balance of the two forces will lead to a long-term average directional motion with constant velocity $\langle z(t) - z(0) \rangle / t = BF$, where $B$ is the mobility. According to the Einstein relation, a temperature independent of $F$ and the Brownian particle can be obtained by $k_B T_{FD} = D/B$. The diffusive and vertical directional motion of HPs in our sheared granular system can be connected by a $T_{FD}$ basically the same way [17,31]. In the following, we denote $t$ the shear cycle number and set $k_B = 1$ for brevity.

We note that the vertical displacements of HPs include both the buoyant motion with respect to the BPs, and a convective flow. This is manifested by the significantly enhanced vertical displacements of the BPs compared with those along the other two directions [Fig. 1(b)].

To obtain the mean relative displacement $\Delta z_{rela}$ of the HPs, we subtract this convective motion of the BPs, that is, their average vertical displacement. The resulting $\Delta z_{rela}$ depends on the time interval $\Delta t$ for all HPs linearly, from which a mobility $B$ can be properly defined:

$$\Delta z_{rela}(\Delta t) = \langle z_{rela}(t_0 + \Delta t) - z_{rela}(t_0) \rangle = BF\Delta t, \qquad (2)$$

where $\langle ... \rangle$ denotes the averages over all HPs and different starting shear cycle number $t_0$, $F = \Delta mg$ is the effective buoyancy force acting on the HPs [31-33]. Figure 2(a) demonstrates that HPs with larger $\Delta m$ have a steeper increasing of $\Delta z_{rela}$ versus $\Delta t$ and the resulting velocity is linear in $\Delta m$ if $\Delta m/m_0 > 0.079$, thus allowing to obtain the mobility $B$. Similarly, the diffusion constant $D$ of HPs can be extracted from their MSD curves:

$$z_{rela}(\Delta t)^2 = \langle [z_{rela}(t_0 + \Delta t) - z_{rela}(t_0)]^2 \rangle = 2D\Delta t. \qquad (3)$$

In Fig. 2(b), the MSD curves show a crossover from the initial sub-diffusion to the long-term normal diffusion and hence we only include data in the diffusive regime to calculate $B$ and $D$ [insets of Figs. 2(a) and 2(b)]. Note that $F$ is estimated to be only about 1% of the average contact force between particles [22], and therefore the buoyancy-induced directional motions of the HPs are too small to modify their MSD behaviors in this experiment. As a result, a temperature $T_{FD}$ based on the Einstein relation can be obtained for $\Delta m/m_0 \geq 0.079$:

$$T_{FD} = \frac{D}{B} = \frac{z_{rela}(\Delta t)^2}{2\Delta z_{rela}(\Delta t)/F}. \qquad (4)$$

In Fig. 3(a), we clearly observe that for all $\Delta m$ the fluctuations and responses collapse on the same linear relationship, the slope of which is just $T_{FD}$. This indicates that the effective temperature obtained by fluctuation-dissipation theorem is a well-defined temperature-like quantity, which reflects the states and characteristics of the driven granular system, irrespective of the various mass difference of the HPs. Moreover, $T_{FD}$ is clearly different for SS and BS systems with the same $\gamma = 0.05$, see Fig. 3(b). This implies that the packing structures and the underlying mechanism of exploring the configurational space of the SS and BS systems are rather different for a same $\gamma$. We also check the dependence of $T_{FD}$ on pressure at $\gamma = 0.05$ by varying the weight of the top lid. We found that, although the resulting disordered packings remain similar, $T_{FD}$ shows a clear linear dependence on $p$, see Fig. 3(b). Note that the ratio $T_{FD}/p$ has the dimension of volume (in unit of $d^3$), which implies that granular packings under cyclic shear relax through free volume [34], reminiscent of compactivity in the Edwards ensemble.

In the Edwards ensemble, the volume fluctuation of a steady-state granular packing can be characterized by a Boltzmann distribution which defines a temperature-like quantity, the compactivity $\chi$. In practice, $\chi$ can be obtained by a histogram overlapping method [20], i.e., by calculating the ratio between probability distribution functions (PDFs) of the local volume in a packing and its reference random loose packing (RLP). As shown in Fig. 4(a), the RLP states of the SS and BS systems are different due to their different friction [20]. The logarithm of the ratio depends linearly on $v$, the slope of which is the compactivity $\chi$ in unit of the particle volume [inset of Fig. 4(a)], and its slope depends on the type of particle. To further examine the quantitative relationship between $T_{FD}$ and $\chi$, we perform additional experiments with different cyclic shear amplitude $\gamma$ under the same free surface condition. For both SS and BS systems, we find $T_{FD}$ varies linearly with $\chi$, with a proportionality factor $T_{FD}/p\chi = 0.91 \pm 0.12$, see Fig. 4(b). We thus draw the conclusion that $T_{FD}/p$ equals the compactivity $\chi$ within the experimental uncertainty.

This equivalence between $T_{FD}$ and the effective temperature defined via the Edwards ensemble has been identified in theoretical calculations of an aging glass with low heat bath temperature [17,35-39]. For both systems, the effective temperatures are closely related to the configurational entropy (or complexity) defined via the local minima (or inherent structures) of the energy landscape [40]. The analogy of these two systems can be justified by the fact that a jammed frictionless granular packing can be mapped onto a local minimum of the free-energy landscape of hard-sphere glasses [41]. The equivalence of two effective temperatures therefore signals that a new type of ergodicity exists when the system enters the glassy landscape regime. This equivalence persists in our frictional granular systems, suggesting that the same ergodicity is preserved despite the complex frictional forces and dissipative interactions between granular particles. From this perspective, many thermodynamic concepts defined on glass systems are still applicable for dense granular materials.

Despite equivalence between $T_{FD}$ and $\chi$ identified above in our system, we point out that there exists a qualitative difference between the microscopic mechanisms of the drag force for the Brownian motion in a granular packing and an ordinary liquid. In the inset of Fig. 2(a), the steady-state relative speed $\Delta z_{rela}/\Delta t$ of the HPs for $\Delta m/m_0 \geq 0.079$ shows a linear relationship with the applied buoyancy force, which is analogous to the viscous behavior of a thermal liquid. However, we also note a distinct behavior of a simple liquid as $v_{rela}$ increases more quickly if $\Delta m/m_0 < 0.079$. We speculate this to be a result of the different microscopic origin of the drag force in a granular "fluid" [31,33]. Unlike particles in thermal fluids in which the viscous drag force is generated by the variation of velocity distribution of the liquid surrounding the HPs, in a quasi-static granular packing, the particles interact through static contact forces. Simply varying the relative contact sliding speeds would leave the frictional force unmodified and therefore it cannot be responsible for the speed-dependent drag force. Instead, we conjecture that the viscous force originates from the asymmetric distribution of contacts on the HP surface. We characterize this asymmetric distribution by the average contact number in the upper and lower hemisphere, $Z_{up}$ and $Z_{low}$, respectively, for both HPs and BPs [28,42], see Fig. 5. The results show that for $\Delta m / m_0 < 0.079$, the contact distributions of HPs and BPs are the same. However, for $\Delta m / m_0 \geq 0.079$, $Z_{up}$ of the HPs starts to increase linearly with the mass difference $\Delta m$ while $Z_{low}$ of the HPs remains at the same value as that of the BPs. We can therefore speculate that an external "effective drag force" is produced as follows: owing to its buoyant tendency, HPs tend to pile up the BPs on their top, hence accumulating a denser "cap" on the upper hemisphere, as shown in the inset of Fig. 5. Assuming that each contact has similar contact and frictional force, this will generate an average "effective drag force" proportional to the asymmetry of contact distribution. This can also naturally explain why there exists a qualitative difference of the viscous behavior when $\Delta m / m_0$ is below or above 0.079. When $\Delta m / m_0$ is small, there exists no significant asymmetry in the contact structure and the viscous drag force is mainly generated by mobilizing the frictional contacts; when $\Delta m / m_0$ is sufficiently large, all the frictional contacts around the HP are mobilized, and the viscous force is then generated by the asymmetric distribution of contact and frictional forces on the particle surfaces.

In summary, using x-ray tomography, we have presented the first experimental evidence that the two types of effective temperatures derived from the FDR and Edwards volume ensemble, coincide with each other in a driven frictional granular system. This finding is robust with respect to different types of hollow particles, cyclic shear amplitudes, and even particle surface roughness. Since granular materials belong to the wide class of disordered materials, the presented validation of the concept of effective temperature consolidates the very foundation of related rheological theories, like shear transformation zone [9] or soft glass rheology [10], in which the effective temperature plays a crucial role in connecting the microscopic or mesoscopic information with the macroscopic plasticity or complex flowing. In a broader sense, our work enlightens the relationship between disordered materials and their transport properties using a statistical mechanics framework for general non-equilibrium systems.

Y. Y. acknowledges support from the fellowship of China Postdoctoral Science Foundation (No. 2021M702151). The work is supported by the National Natural Science Foundation of China (No. 11974240), the Science and Technology Innovation Foundation of Shanghai Jiao Tong University (No. 21X010200829), and the Science and Technology Commission of Shanghai Municipality (No. 22YF1419900). W. K. is a senior member of the Insitut Universitaire de France.

Corresponding author

[*]yuanyeoct20@sjtu.edu.cn

[†]yujiewang@sjtu.edu.cn

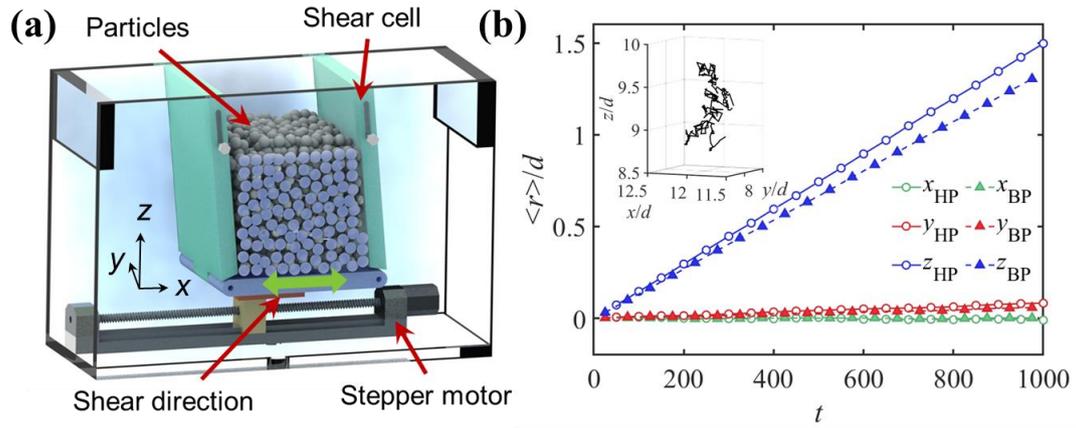

FIG. 1. (a) Schematic of the experimental setup. (b) Average displacements $\langle r \rangle$ in three directions of HPs and BPs as a function of $t$. Inset: trajectory (length $t=1000$) of a HP under cyclic shear. Results in (b) are for the SS systems with free top surface, $\gamma = 0.05$, and $\Delta m / m_0 = 0.376$.

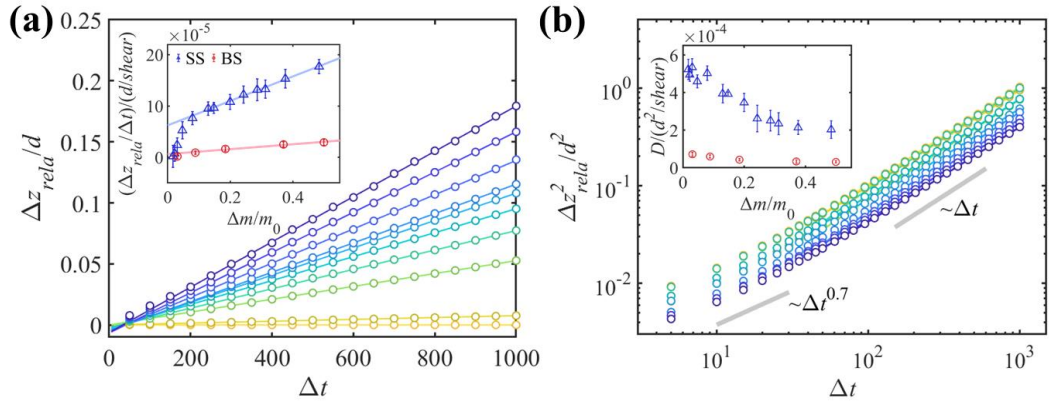

FIG. 2. (a) Relative vertical displacements $\Delta z_{rela}$ as a function of $\Delta t$ with $\Delta m / m_0 = 0.484$, 0.376, 0.312, 0.243, 0.200, 0.130, 0.079, 0.047, 0.022 and 0 (from top to bottom). The solid lines denote the linear fits to the data. Inset: relative vertical speeds $\Delta z_{rela} / \Delta t$ as a function of $\Delta m / m_0$. The solid lines are the linear fits to $v_{rela}$ for $\Delta m / m_0 \geq 0.079$. (b) MSDs for different $\Delta m/m_0$ as a function of $\Delta t$, where sub-diffusion and normal diffusion regions are marked by the two solid lines. Inset: self-diffusion coefficient $D$ as a function of $\Delta m / m_0$. Results in (a) and (b) are for the HPs in the SS systems with free top surface and $\gamma = 0.05$.

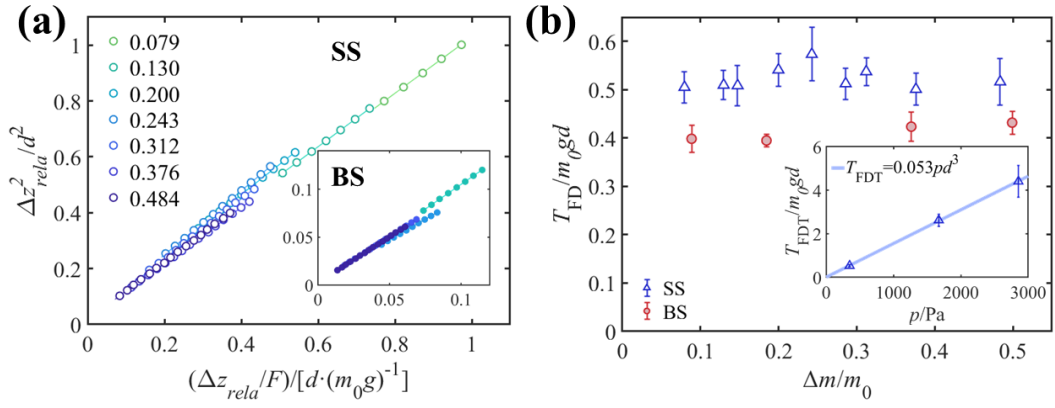

FIG. 3. (a) Parametric plot of diffusion versus mobility for the SS systems and different $\Delta m/m_0$ (legend). Inset shows the same parametric plot for the BS systems. (b) $T_{FD}$ for SS and BS systems as a function of $\Delta m/m_0$. Inset: $T_{FD}$ as a function of pressure $p$ for the SS system. Results in (a) and (b) are for systems with $\gamma = 0.05$.

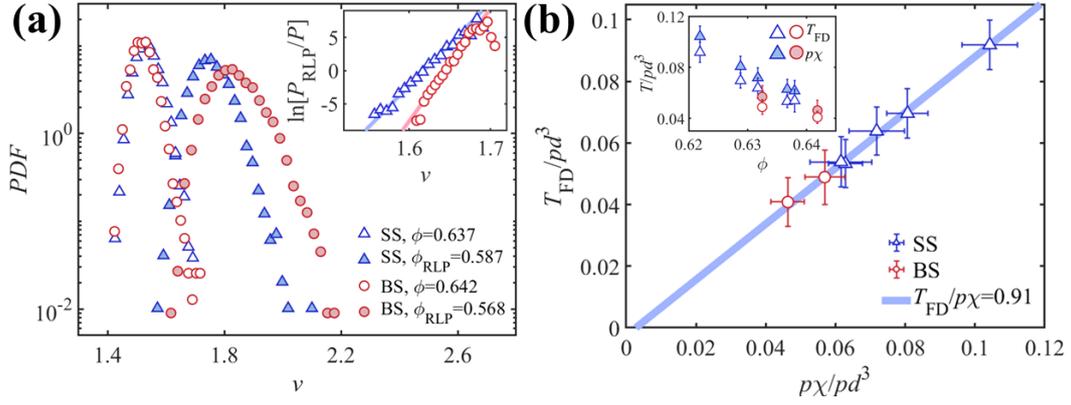

FIG. 4. (a) PDFs of local volumes $v$ for SS and BS packings under cyclic shear ($\gamma = 0.05$) and their respective reference RLP states. Their global packing fractions are included in the legend. Inset: the ratio of PDFs of $v$ between sheared packings and their respective RLP states. (b) Effective temperature calculated via the FDR versus Edwards' compactivity for SS and BS systems. Solid line is the linear fit: $p\chi/T_{FDT} = 1.1 \pm 0.15$. Inset: $T_{FDT}$ and $p\chi$ at different $\phi$, tuned by $\gamma = 0.03$, 0.05, 0.08, 0.12, 0.20 for the SS systems and $\gamma = 0.03$, 0.012 for the BS systems.

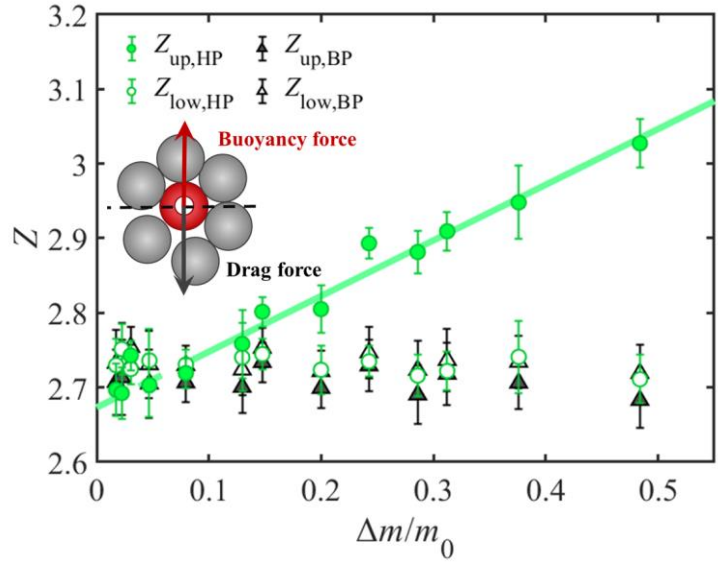

FIG. 5. The average contact number in the upper and lower hemisphere, $Z_{up}$ and $Z_{low}$, as a function of $\Delta m / m_0$ for both HPs and BPs in the SS systems with free top surface. Solid line is the linear fit to $Z_{up}$ of HPs for $\Delta m / m_0 \geq 0.079$. Inset is a schematic diagram describing how the viscous-like drag force on the HPs (red) is generated by the asymmetric distribution of the BPs (gray) in direct contact.

# Supplemental Materials for
# Equivalence of Fluctuation-Dissipation and Edwards' Temperature in Cyclically Sheared Granular Systems

**Measurement of pressure**

In order to accurately measure the pressure on particles in the packing, we modify the original shear box, as shown in Fig. S1(a). Specifically, the bottom plate of the shear cell consists of two independent parts, i.e., the disc and the plate, with a 2-mm gap to avoid the influence of friction, see Fig. S1(b). A force probe (sensitivity 0.1 N, eDPU-200N, IMADA Company) is fixed under the disc via a vertical metal rod in order to measure the pressure directly applied by the particles on the disc.

We first perform a calibration by filling up the shear cell with a plastic bag of water. Changing the height of water, e.g., $h$=10 cm, we measure the pressure $p = F/S = 980.4$ Pa where $F = 5.0 \pm 0.1$ N is the force measurement and $S = 0.0051$ m$^2$ is plate area. The result is consistent with the standard water pressure $p = \rho g h$ where $\rho$ is the density of water.

Then, we perform experiments of granular pressure measurement. The internal granular pressure at a certain depth $H$ from the top is identical to the bottom pressure of a granular packing with the same height $H$. The latter is exactly accessible here. According to the main text, we prepare the packings by applying hundreds of shear cycles, with different heights $H$ and upper boundary conditions (free or covered by a lid with mass $M = 1.95$ or 3.85 kg). After that, we replace the original horizontally placed step motor with the force-measuring component. Figure S1(c) shows the measured bottom pressure as a function of $H$ for three upper boundary conditions. Considering that the average depth of hollow particles (HP) is about

50±10 mm during the temperature measurements, the pressure on the HPs are $348\pm70$, $1672\pm104$ and $2855\pm174$ Pa for $M=0$, 1.95 and 3.85 kg, respectively.

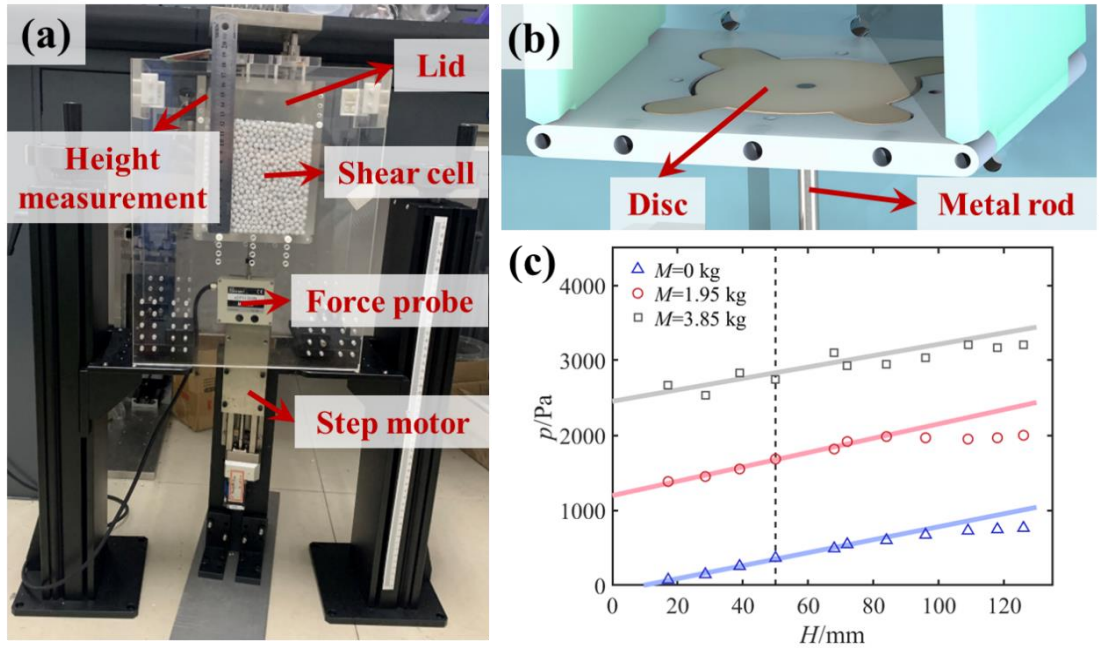

FIG. S1 (a) The modified experimental setup of the granular pressure measurement. (b) Schematic of the bottom disc for force measurement. (c) The pressure $p$ at the bottom of the granular packings as a function of height $H$ for three upper boundary conditions. Dotted line denotes the average depth of hollow particles (HP) $H$=50 mm.